\documentclass[prb,twocolumn,showpacs,amsmath,amssymb,groupedaddress]{revtex4-1}
\usepackage{graphicx}
\usepackage{color}

\def\be{\begin{equation}}
\def\ee{\end{equation}}
\def\ba{\begin{eqnarray}}
\def\ea{\end{eqnarray}}
\def\bc{\begin{center}}
\def\ec{\end{center}}
\def\p{\partial}
\def\sgn{{\rm sgn}}

\begin{document}
\title{Radiative damping and synchronization in a graphene-based terahertz emitter}

\author{A. S. Moskalenko}
\email[Email: ]{andrey.moskalenko@physik.uni-augsburg.de}
\altaffiliation[Also at ]{Ioffe Physical-Technical Institute of
RAS, 194021 St.~Petersburg, Russia}

\affiliation{Institute of Physics, University of Augsburg, D-86135
 Augsburg, Germany}

\author{S. A. Mikhailov}
\email[Email: ]{sergey.mikhailov@physik.uni-augsburg.de}
\affiliation{Institute of Physics, University of Augsburg, D-86135 Augsburg, Germany}

\date{\today}

\begin{abstract}
We investigate the collective electron dynamics in a recently
proposed graphene-based terahertz emitter under the influence of
the radiative damping effect, which is included self-consistently
in a molecular dynamics approach. We show that under appropriate
conditions synchronization of the dynamics of single electrons
takes place, leading to a rise of the oscillating component of the
charge current. The synchronization time depends dramatically on
the applied dc electric field and electron scattering rate, and is
roughly inversely proportional to the radiative damping rate that
is determined by the carrier concentration and the geometrical
parameters of the device. The emission spectra in the synchronized
state, determined by the oscillating current component, are
analyzed. The effective generation of higher harmonics for large
values of the radiative damping strength is demonstrated.
\end{abstract}

\pacs{78.67.Wj (Optical properties of graphene), 05.45.Xt
(Synchronization, nonlinear dynamics), 41.60.-m (Radiation by
moving charges)}

\maketitle
\section{Introduction}
The electromagnetic radiation of terahertz frequencies
($0.1\lesssim \nu\lesssim 10$ THz) has many potential applications
in medical imaging, security, astronomy and other areas. The
sources of coherent and powerful THz radiation are, however,
mainly restricted to vacuum devices like backward wave oscillators
or free electron lasers. The operation principle of these devices
is based on the Smith-Purcell effect \cite{Smith1953}:
electromagnetic waves are emitted by a fast electron beam moving
across a periodic potential. The radiation frequency of the
Smith-Purcell emission,
 \begin{equation}
   \nu_{_{\rm SP}}=\frac{v_{0}}{a_x}\label{smith},
 \end{equation}
is determined by the drift velocity of the electrons $v_{0}$ and
the period of the potential $a_x$. It can be controlled by a dc
voltage which accelerates electrons up to the velocity $v_{0}$.

The vacuum devices are bulky and expensive. The need for a compact
source of terahertz radiation stimulated experiments (e.g.
\cite{Tsui80,Hirakawa95,Otsuji11}) attempting to create a
solid-state Smith-Purcell emitter. In such experiments
semiconductor structures, like GaAs/AlGaAs quantum wells, have
been used, with a metallic grating evaporated on top of the
system. Electrons were driven in a two-dimensional (2D) channel in
the direction perpendicular to the grating stripes, but instead of
the coherent Smith-Purcell emission with the velocity-dependent
frequency (\ref{smith}) a weak thermal radiation at the frequency
of 2D plasmons $\nu_{\rm p}=\omega_{\rm p}/2\pi$ was observed. The
reason for this failure was clarified in
Ref.~\onlinecite{Mikhailov1998}. It was shown that the strong
coherent radiation is observed at the frequency (\ref{smith}) only
if the drift velocity $v_{0}$ substantially exceeds a threshold
value \be v_0\gg v_{\rm th} \simeq \nu_{\rm p}a_x =
\sqrt{\frac{n_{\rm s}e^2a_x}{m\epsilon}},\label{thresh} \ee where
$e$, $m$ and $n_{\rm s}$ are the charge, the effective mass and
the 2D density of electrons and $\epsilon$ is the dielectric
permittivity of the surrounding medium. In vacuum devices the
condition (\ref{thresh}) is easily satisfied, but in
semiconductors the plasma frequency is very large so that
typically the relation $v_0\ll v_{\rm th}$ holds. In this case,
however, the system emits at the plasma frequency $\nu_{\rm p}$
due to the heating of the electron gas (the thermal radiation). As
seen from Eq. (\ref{thresh}) the Smith-Purcell emission can be
realized if a small amount of electrons can be driven with a
sufficiently large velocity.

\begin{figure}
  \includegraphics[width=8.4cm]{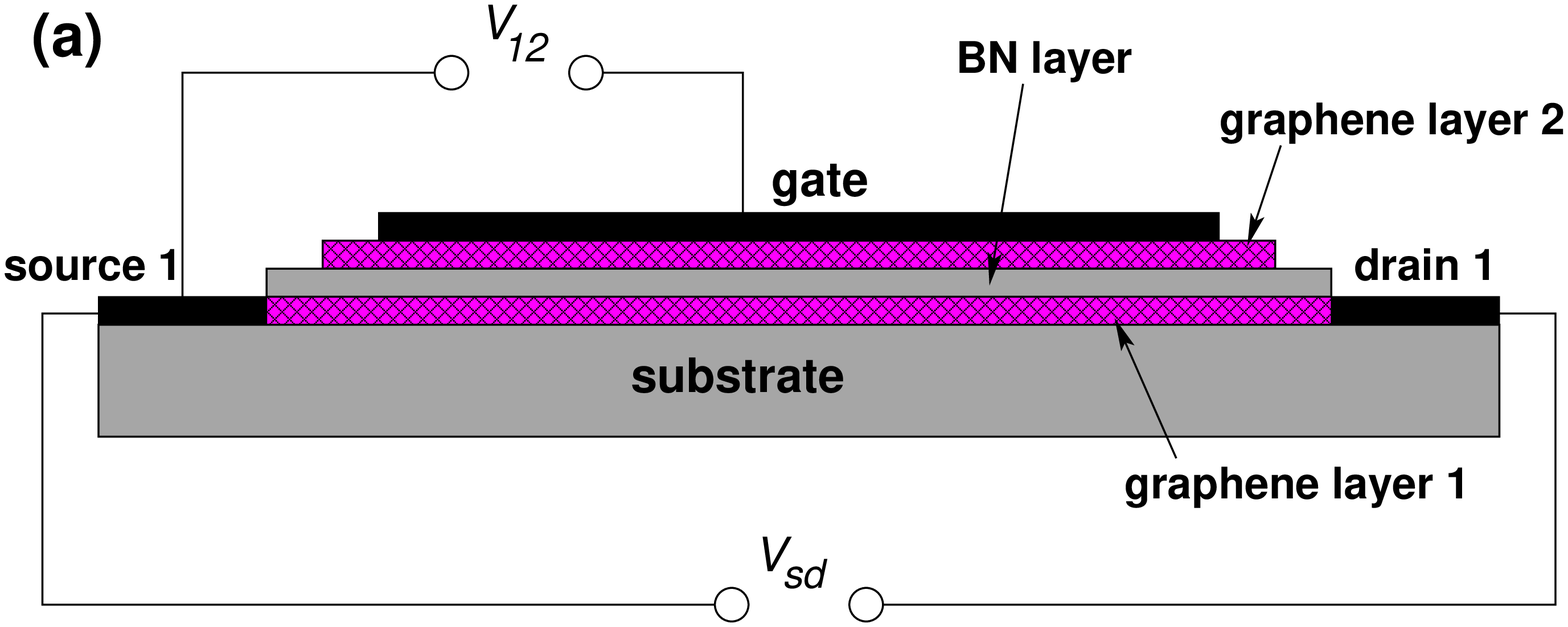}\\
\includegraphics[width=4.2cm]{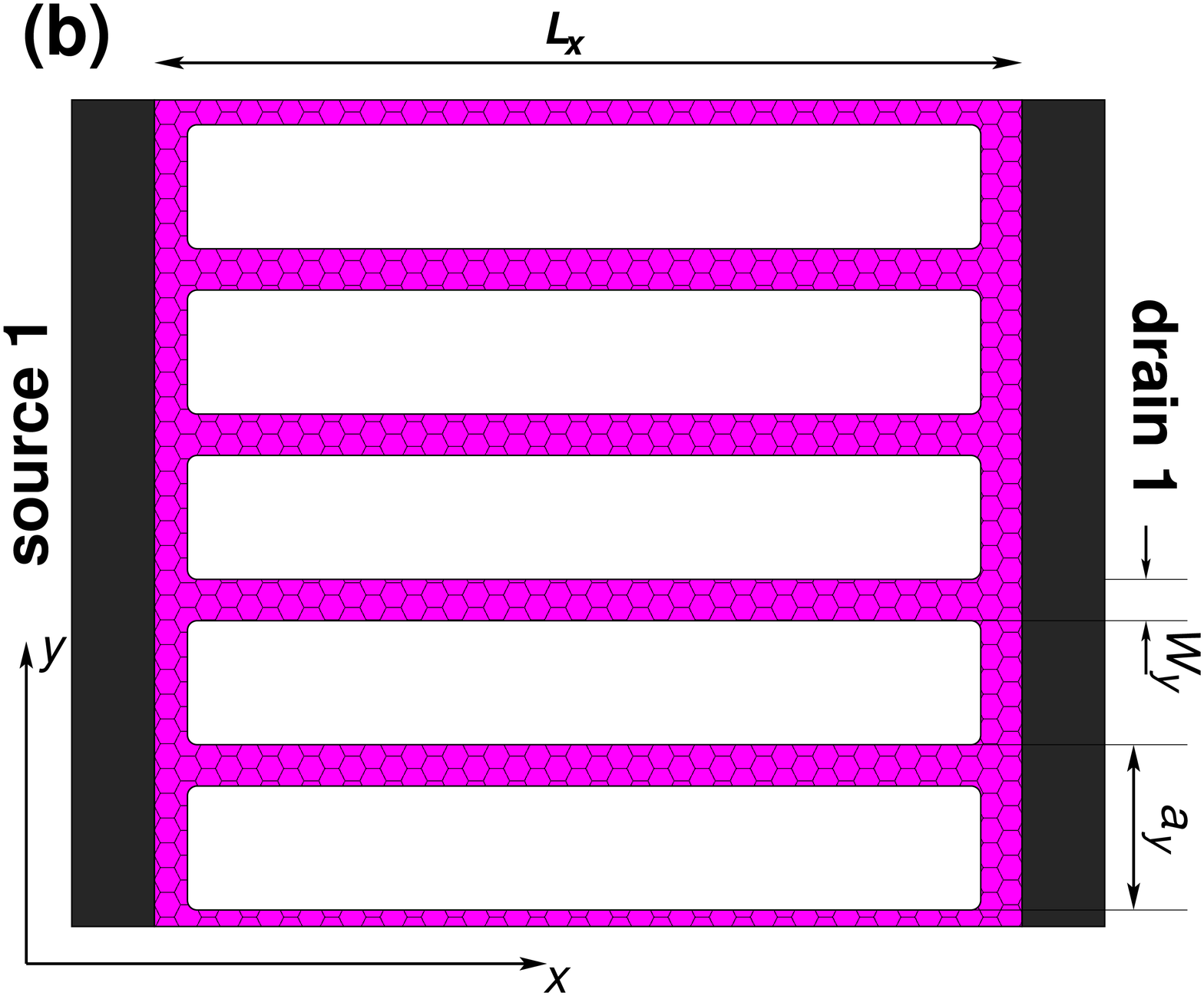}
\includegraphics[width=4.0cm]{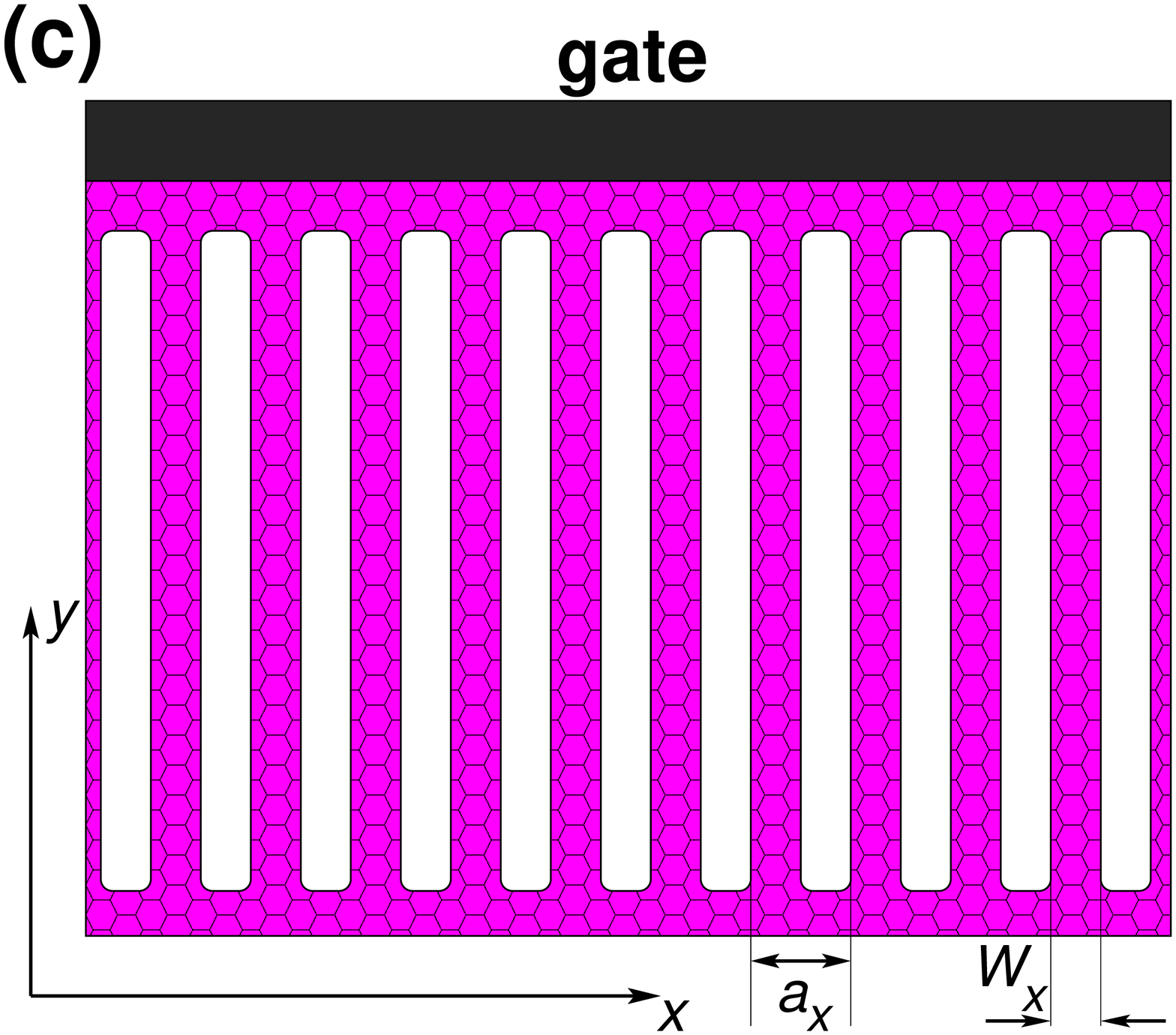}
  \caption{(Color online) The geometry of the device: (a) the side view, (b) the top view of the first (active) graphene layer, (c) the top view of the second graphene layer. The first graphene layer lying on
  the dielectric substrate has the form of an array of stripes with the width $W_y$ and the period $a_y$.
  The second layer (the gate) serves as a grating with the stripe width $W_x$ and the period $a_x$.
   The layers are separated by a thin dielectric (e.g. BN) of thickness $d$.
   In the operation mode a large (driving) dc voltage $V_{\rm sd}$ is applied between the source and drain and
   a weak dc voltage $V_{12}$ creating the periodic potential (\ref{Eq:U_x}) is applied between the gate and the first graphene layer.
   The length of the sample in the $x$-direction is $L$.
           \label{geom}}
\end{figure}

The discovery of graphene \cite{Novoselov04,Novoselov05,Zhang05}
opened great opportunities in exploring unique properties of this
material in scientific research and practical applications.
Recently it was shown \cite{Mikhailov2013} that, due to the
massless energy dispersion and the very large Fermi velocity of
graphene electrons (as compared to semiconductors), the
Smith-Purcell emission condition (\ref{thresh}) can be realized in
graphene. A specific device structure proposed in
Ref.~\onlinecite{Mikhailov2013} consists of two graphene layers
lying on a substrate and separated by a thin boron-nitride
dielectric (see Fig.~\ref{geom}). The first graphene layer
(Fig.~\ref{geom}b) has the form of an array of narrow stripes
oriented along the $x$-axis. The second graphene layer (the gate,
Fig.~\ref{geom}c) consists of a set of perpendicular stripes with
the period $a_x$ and serves as a grating coupler. A strong dc
current is driven in the first (active) layer from the left
(source) to the right (drain) contact, and the periodic potential
is produced by applying a weak dc voltage between the two graphene
layers. The calculations of Ref.~\onlinecite{Mikhailov2013} showed
that the proposed structure should be able to coherently emit at
frequencies ranging from fractions of terahertz up to $\simeq 30$
THz with the power density up to 0.5 W/cm$^2$.

The theory of Ref.~\onlinecite{Mikhailov2013} is based on a
single-particle approach which corresponds to the cold plasma
approximation. In such an approach one considers the motion of a
single particle under the action of the driving dc electric field
and the periodic potential, and ignores the thermal (or Fermi)
distribution of electrons over quantum states. In this paper we
treat the problem within the molecular dynamics approach, taking
into account the electron statistics as well as the back action of
the emitted radiation
to the electron dynamics in the structure. Starting from a
homogeneous thermal distribution at the applied dc electric field,
we analyze the nonlinear dynamics of the electron momentum and
coordinate distributions, as well as the resulting current density
dynamics, after the periodic potential is switched on. We
demonstrate how the back action of the radiation synchronizes the
oscillatory motion of electrons and stimulates all electrons to
emit coherently. The synchronization of the electron motion (the
radiative damping) is a prerequisite for an effective emission
from the device, therefore the understanding of the studied
processes is important for the efficient device operation.

The synchronization itself is an interesting phenomenon in
nonlinear science which was widely studied in physical, chemical,
biological and social systems
\cite{Strogatz2001,Pikovsky2003,Arenas2008}. Typically, the
synchronization occurs on a much longer time scale than the
characteristic time scales of single oscillators. The onset of the
synchronization and the dynamics of the order parameters
characterizing the synchronization belong to open general problems
in nonlinear science \cite{Strogatz2000}. In this work we address
these problems numerically in the particular case of the electron
dynamics in the proposed graphene-based terahertz emitter.

\section{Theoretical model}

\begin{figure}[t!]\centering
  \includegraphics[width=\columnwidth]{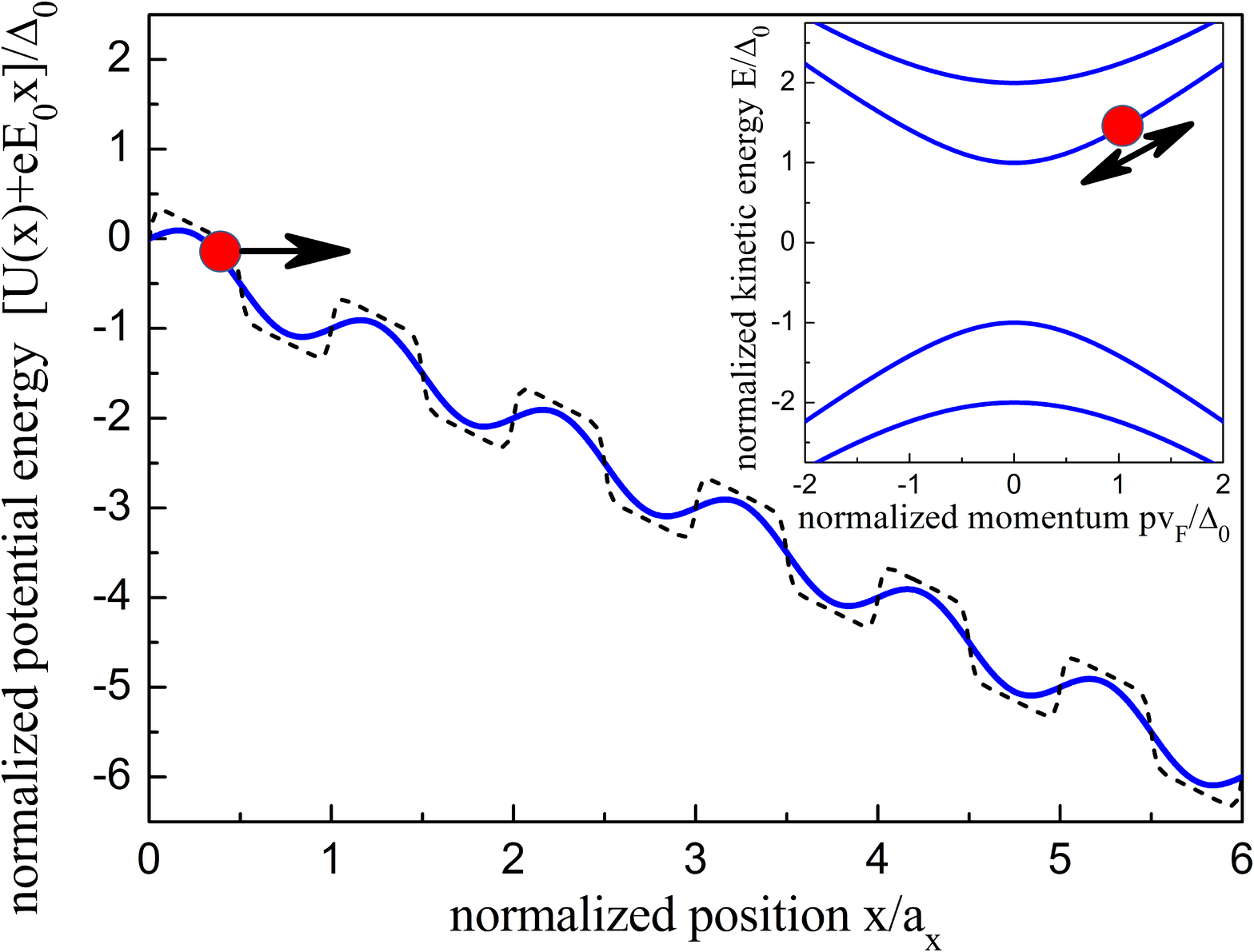}
  \caption{(Color online) Electron moving in a total potential  $U(x)+eE_0x$ created by the periodic potential
   (\ref{Eq:U_x}) and the dc driving field $E_0$. For this illustration we selected the value of the applied dc electric
  field such that $-eE_0a_x/\Delta_0=1$ whereas
  $U(x)$ is determined by Eq.~\eqref{Eq:U_x} with $S=0.1$ and $U_0/\Delta_0=0.3$ (solid blue line) and $S=5$ and $U_0/\Delta_0=0.4$
  (dashed black line). Inset
  illustrates the energy band structure of a graphene stripe and the
  variations of the electron kinetic energy in the first
  conduction subband.
           \label{Fig:system}}
\end{figure}

Following Ref.~\onlinecite{Mikhailov2013} we consider the motion
of electrons in graphene stripes of the first layer (Fig.
\ref{geom}) under the action of the dc electric field $E_0=V_{\rm
sd}/L$ and the periodic potential $U(x)$ created by the voltage
$V_{12}$ between the layers. The spectrum of electrons in the
stripes has the form $E_{\pm,n}(p)=\pm \sqrt{\Delta_0^2
n^2+v_{_{\rm F}}^2p^2}$ (see inset to Fig.~\ref{Fig:system}),
where $n=1,2,\ldots$. The gap at the Dirac point $2\Delta_0$ is
assured by an appropriate choice of the stripe boundary conditions
and is of the order of  $\pi\hslash v_{_{\rm F}}/W_y$
\cite{Brey2006,Yang2007,Akhmerov2008,Mikhailov2013}, where
$v_{_{\rm F}}$ is the Fermi velocity in the bulk graphene. The
periodic potential seen by electrons of the main layer is modeled
as
\begin{equation}\label{Eq:U_x}
  U(x)=U_0\frac{\tanh\left[S\sin(2\pi x/a_x)\right]}{\tanh S}\;,
\end{equation}
where $U_0$ and $a_x$ are the amplitude and the period of the
potential, and the parameter $S$ characterizes its steepness: at
$S\ll 1$ the potential is practically sinusoidal, whereas at $S\gg
1$ it has a periodic rectangular shape and contains many spatial
Fourier harmonics. In the experiment, the strongly non-sinusoidal
regime $S\gg 1$ is realized if the distance $d$ between the main
and the grating graphene layers is much smaller than the grating
period $a_x$ (for example, if the BN dielectric layer between the
graphene sheets is about $5-10$ nm while the grating period
exceeds $0.1-1$ $\mu$m). The total potential seen by the electrons
in the graphene stripes is shown in Fig.~\ref{Fig:system}.

In order to model the dynamics of electrons  we apply a molecular
dynamics approach and describe it by the following equations of motions:
\begin{eqnarray}
\frac{{\rm d}{x_i}}{{\rm d}t}&=&v_i=\frac{v_{_{\rm
F}}^2p_i}{\sqrt{\Delta_0^2+v_{_{\rm F}}^2p_i^2}} \;, \ \ \
i=1,2,\dots,N, \label{Eq:xi}\\
\frac{{\rm d}{p_{i}}}{{\rm d}t}&=&-\gamma p_{i}-e  E_0 -\frac{{\rm d}U(x_i)}{{\rm d}x_i}
   -eE_{\rm ac}(t). 
\label{Eq:pi}
\end{eqnarray}
Here $-e$, $x_i$, $p_i$ and $v_i$  are the charge, coordinate,
momentum and velocity of the $i$-th electron, respectively,
whereby the total number of electrons amounts to $N$. The first
term on the right hand side of Eq.~\eqref{Eq:pi} accounts for the
relaxation processes, whereby $\gamma$ is the phenomenological
scattering rate. The physical meaning of the last term in
Eq.~\eqref{Eq:pi} can be explain as follows. If the periodic
potential is absent and electrons move only under the action of
the dc electric field $E_0$, their velocity $v_i$ and the total
current
\begin{equation}\label{Eq:j}
  j_x=-en_{\rm s}\frac{1}{N}\sum_{i=1}^N v_i=-en_{\rm s}\frac{1}{N}\sum_{i=1}^N
  \frac{v_{_{\rm F}}^2p_i}{\sqrt{\Delta_0^2+v_{_{\rm F}}^2p_i^2}}\;,
\end{equation}
are time-independent, and the ac electric field $E_{\rm ac}(t)$ is
zero. Here $n_{\rm s}$ denotes the average two-dimensional
electron density. If the potential $U(x)$ is switched on,
electrons moving across this potential with a large average
velocity $v_0$ begin to oscillate in time, and hence, to emit
electromagnetic waves. The electric $E_{x}(z,t)$ and magnetic
$H_y(z,t)$ fields of this wave are proportional to
$e^{i\omega|z|/c-i\omega t}$, where $c$ is the light velocity and
$\omega$ is the frequency of the emitted radiation. The force
$-eE_{\rm ac}(t)$ in Eq.~\eqref{Eq:pi} is due to this field at
$z=0$, $E_{\rm ac}(t)\equiv E_x(0,t)$; it describes the back
influence of the radiation on the electron dynamics and works as
an additional friction force (the radiative damping).

The field $E_{\rm ac}(t)$ is related to the current density
(\ref{Eq:j}) by the Maxwell equations. In Appendix
\ref{App:rad_damp} we show that $E_{\rm ac}(t)$ can be written as
\begin{equation}\label{Eq:E_x_result}
  E_{\rm ac}(t)=-\frac{2\pi}{c}
 \left[ j_x(t)-\bar{j}_x(t)\right],
\end{equation}
where
\begin{equation}
\label{Eq:j_bar}
  \bar{j}_x(t)=\gamma_L\int_{-\infty}^t j_x(t')\exp[-\gamma_L(t-t')]\;{\rm d}t',
\end{equation}
and the cut-off parameter
\begin{equation}\label{Eq:omega_L}
  \gamma_L=\frac{2\pi c}{L}
\end{equation}
is related to the length of the sample. The correction
$\bar{j}_x(t)$ ensures that low frequency components of $E_{\rm
ac}$ with $\omega\ll \gamma_L$ vanish because in this case the
radiation wavelength $\lambda$ is much larger than the length of
the sample $L$. At high frequencies $\omega\gg \gamma_L$ the
correction $\bar{j}_x(t)$ reduces to an average dc current at the
time moment
$t$, which does not contribute to the emission and therefore to the radiative damping. 
Up to this shift the term $-eE_{\rm ac}$ in Eq.~\eqref{Eq:pi} is
then
of the order $-\Gamma \sum_{i=1}^Np_i/N$, where $\Gamma$ is the
characteristic radiative-damping rate
\begin{equation}\label{Eq:radiative_rate}
  \Gamma=\frac{2\pi n_{\rm s}e^2v_{_{\rm F}}^2}{\Delta_0 c}.
\end{equation}
This model for the radiative damping term $E_{\rm ac}$ is well
applicable and is insensitive to the choice of $L$ if the
condition $2\pi\nu_0\gg\gamma_L$ is fulfilled, where
\begin{equation}\label{Eq:characteristic_frequency}
  \nu_0=v_{_{\rm F}}/a_x.
\end{equation}
Using Eq.~\eqref{Eq:omega_L}  we get then a restriction for the
sample length $L\gg \lambda_0$, where $\lambda_0$ is the
characteristic radiation wavelength. For the grating period
$a_x\sim0.5-5~\mu$m we have $\nu_0\sim 1-10$~THz ($\lambda_0\sim
0.3-0.03$~mm) so that the sample should be longer than $\sim
1-0.1$~mm.

The system of equations to be solved is thus given by Eqs.
(\ref{Eq:xi})--(\ref{Eq:j_bar}). Notice that the radiative damping
term $-eE_{\rm ac}$, proportional to the current of all particles
(\ref{Eq:j}), leads to a coupling of the otherwise uncoupled
single-particle equations (\ref{Eq:xi})--(\ref{Eq:pi}) for $\dot
r_i$ and $\dot p_i$. This leads to a synchronization of the
electron dynamics: while just after the potential $U(x)$ is
switched on all electrons oscillate with different phases
(incoherent emission), after a while their phases becomes
synchronized due to the radiative damping coupling, and the
emission becomes coherent.  As will be seen below, the
synchronization time $\tau_{\rm sync}$ is governed by the ratio
$\Gamma/\nu_0$.
The typical value of $\Gamma/2\pi$ calculated from
Eq.~\eqref{Eq:radiative_rate} lies between 0.1 and 1 THz
\cite{Mikhailov08a} (for comparison, as mentioned, $\nu_0$ is in
the range 1-10~THz). It can be, however, increased significantly
if the graphene emitter is placed in a cavity (e.g. between THz
mirrors \cite{Krumbholz2006}).


\section{Simulation results}
In simulations presented below we fixed the periodic potential
amplitude $U_0$ to $U_0/\Delta_0=0.3$ and assume that only the
lowest electron energy band is populated. The calculations were
performed with $N=10^6$ electrons; the further increase of $N$
practically did not change the results. The cut-off parameter was
chosen to be $\gamma_L=\nu_0$ which corresponds to the device
length $L\approx 2\times 10^3 a_x$, but the results were found to
be insensitive to $\gamma_L$ if it  was varied from $~0.1\nu_0$ to
$~2\nu_0$.

\begin{figure}[t!]\centering
  \includegraphics[width=\columnwidth]{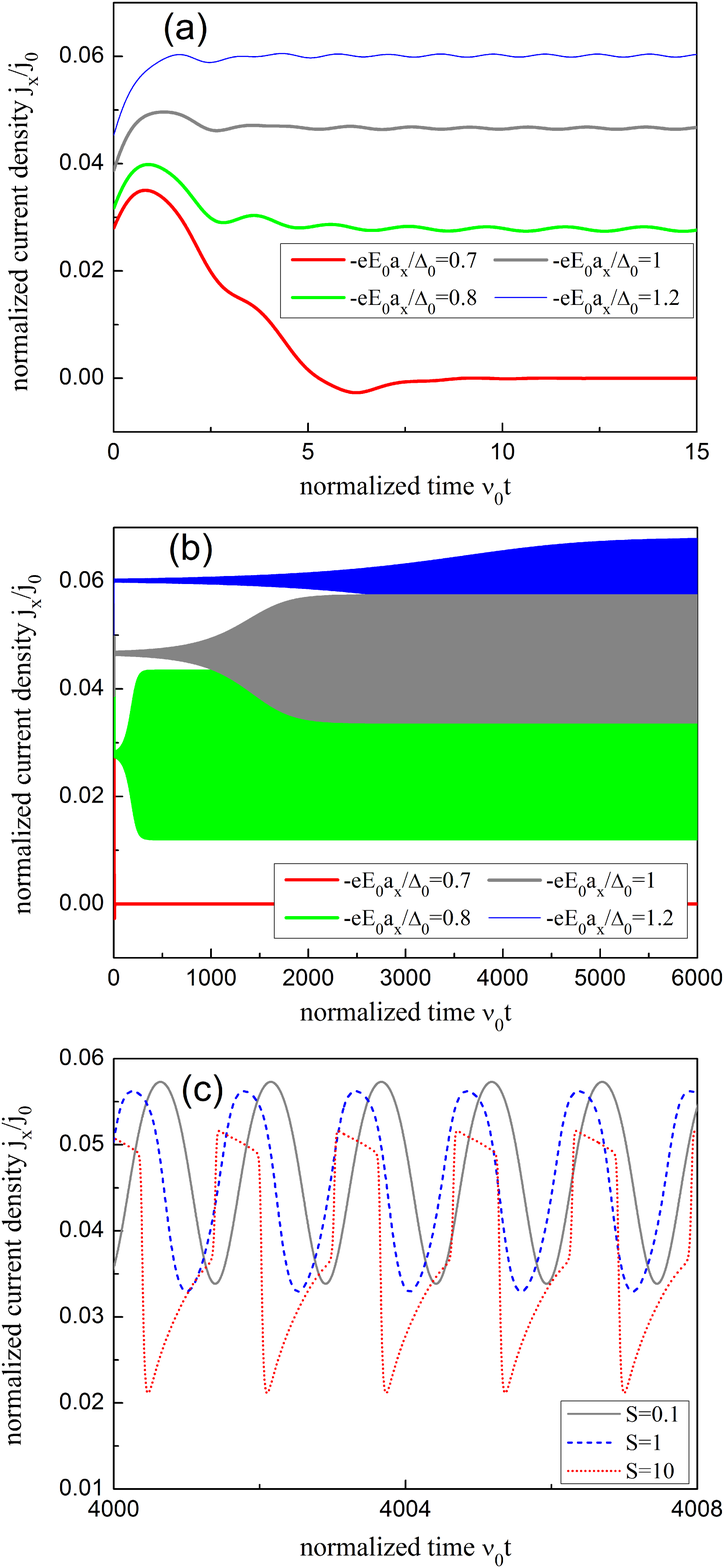}
  \caption{(Color online) (a) Time dependence of the charge current density at a short time scale
           for different values of the applied dc electric field $E_0$.
           Values of $E_0$ shown in the legend increase for the curves from the bottom to the
           top of the figure.
           Other parameters of the simulations: $S=0.1$, $\Gamma/\nu_0=0.1$, and $\gamma/\nu_0=1$.
           The current density is normalized by $j_0=-en_{\rm s}v_{_{\rm F}}$.
           (b) The same dynamics is shown at a long time scale.
           (c) Dependence of the current density oscillations on the periodic potential
           steepness
            parameter $S$. The oscillations are shown in the correspondingly short time interval
            when they have stabilized. Other parameters of the simulations:
           $-eE_0a_x/\Delta_0=1$, $\Gamma/\nu_0=0.1$, and $\gamma/\nu_0=1$.
           \label{Fig:E_S_current}}
\end{figure}

Figure \ref{Fig:E_S_current}a shows the time dynamics of the
generated charge current at different values of the applied dc
electric field $E_0$. Initially, at $t<0$, the periodic grating
potential is switched off. The electrons move in the potential of
the dc field. They have a homogeneous spatial distribution whereas
their concentration is low enough so that their momenta obey a
Boltzmann distribution at temperature $T=1.2\Delta_0$ shifted by
the applied dc electric field (the condition $T=1.2\Delta_0$
corresponds to $T=300$~K for $W_y=0.1~\mu$m). At $t=0$ the
periodic grating potential is switched on. At first, this leads to
a quick redistribution of the electrons, which can be observed in
the charge current dynamics as oscillations relaxing at the time
scale of the order of $1/\gamma$ (see
Fig.~\ref{Fig:E_S_current}a).
The charge current dynamics at a longer time scale is presented in
Fig.~\ref{Fig:E_S_current}b. We see that for small values of $E_0$
the system arrives, after the short period of relaxation, to a
state with no flowing current. An increase of the dc electric
field leads to a change in the current dynamics:
a dc current, along with an initially small
ac current component, appears. The oscillation amplitude of the ac
contribution increases with time and finally saturates.
The
stabilization time of the current dynamics is much longer than the
initial relaxation time and the period of the current
oscillations; therefore they are not resolved in
Fig.~\ref{Fig:E_S_current}b.



The stabilization time strongly depends on the value of the
applied dc electric field. As seen from
Fig.~\ref{Fig:E_S_current}b, a relatively small increase of this
value leads to a dramatic growth of the time required for the
stabilization of the current oscillations. The increase of the dc
electric field leads also to an increase of the dc current
component. On the contrary, the amplitude of the stabilized ac
current oscillations moderately decreases. These oscillations are
resolved in Fig.~\ref{Fig:E_S_current}c, where they are shown at
different values of the steepness parameter $S$. As it was already
discussed in Ref.~\onlinecite{Mikhailov2013}, for small values of
$S$, i.e. $S\ll 1$, the oscillations are essentially harmonic
whereas for $S\gg 1$ the anharmonicity becomes strongly
pronounced.

\begin{figure*}[t!]\centering
  \includegraphics[width=12.0cm]{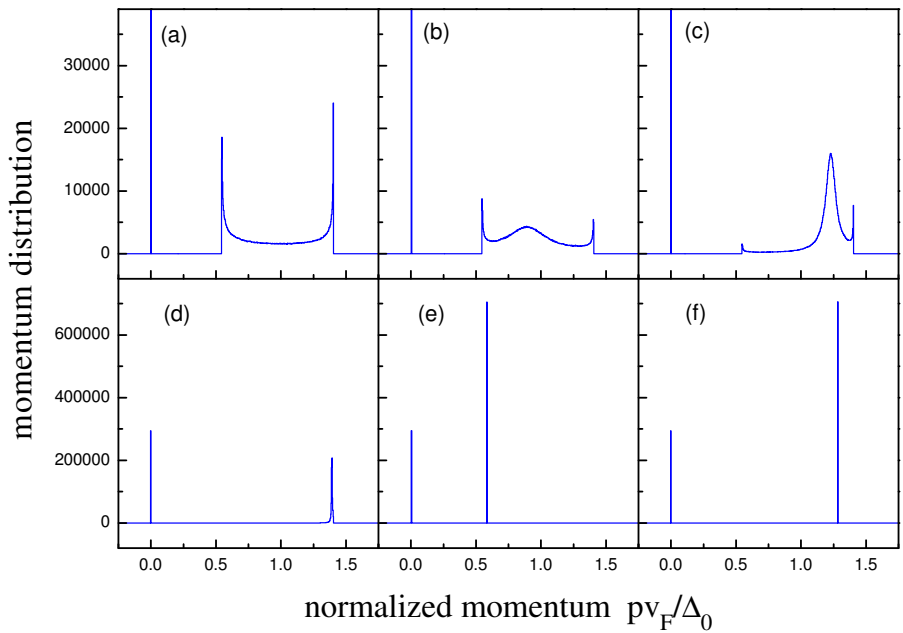}
  \caption{(Color online) Snapshots of the momentum distribution at different time moments: (a) $\nu_0t=200$,
   (b) $\nu_0t=1500$, (c) $\nu_0t=1800$,
  (d) $\nu_0t=2100$, (e) $\nu_0t=4000$, and (f) $\nu_0t=5000$.
  Parameters of the simulation are the same as for the gray curve in Fig.~\ref{Fig:E_S_current}:
  $-eE_0a_x/\Delta_0=1$, $S=0.1$, $\Gamma/\nu_0=0.1$,  and $\gamma/\nu_0=1$.
           \label{Fig:distribution_p_time}}
\end{figure*}

\begin{figure*}[t!]\centering
  \includegraphics[width=12.0cm]{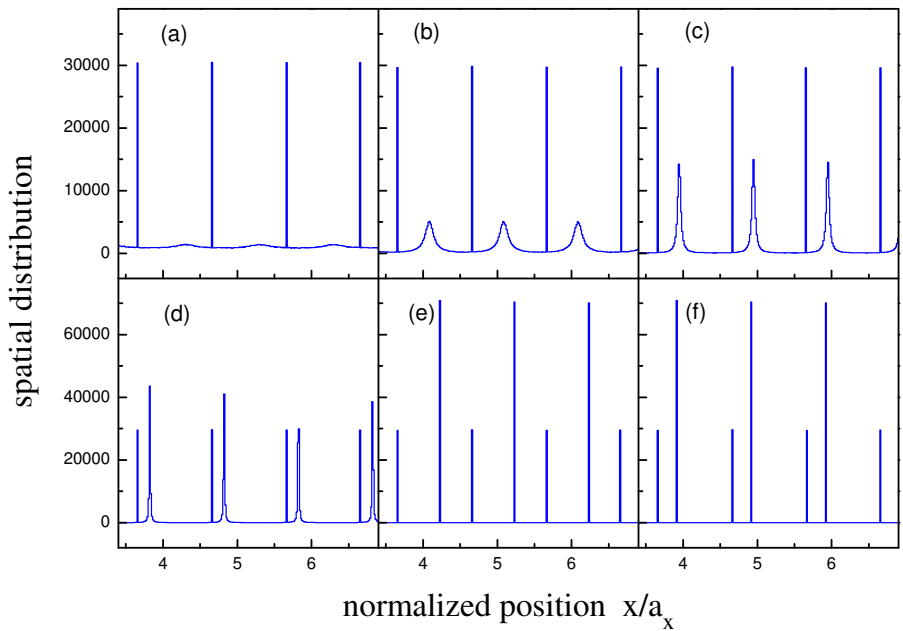}
  \caption{(Color online) Snapshots of the spatial distribution at different time moments: (a)
  $\nu_0t=200$,
  (b) $\nu_0t=1500$, (c) $\nu_0t=1800$,
  (d) $\nu_0t=2100$, (e) $\nu_0t=4000$, and (f) $\nu_0t=5000$.
  Parameters of the simulation are the same as in Fig.~\ref{Fig:distribution_p_time}.
           \label{Fig:distribution_x_time}}
\end{figure*}

In order to understand the observed behavior of the current, and
specifically the rise of its oscillations in time with the
subsequent stabilization, we have analyzed the dynamics of the
coordinate and momentum distributions in the system. The snapshots
of the momentum distribution, corresponding to the current
dynamics shown in Fig.~\ref{Fig:E_S_current}b for
$-eE_0a_x/\Delta_0=1$ (gray curve in Fig.~\ref{Fig:E_S_current}b),
are presented in Fig.~\ref{Fig:distribution_p_time}. The snapshots
of the spatial distribution for the same simulation are shown in
Fig.~\ref{Fig:distribution_x_time}. The first snapshots at $\nu_0
t=200$ are taken already well after the short initial stage of the
current relaxation. From the momentum distribution
(Fig.~\ref{Fig:distribution_p_time}a), we see that a part of the
electrons stopped whereas another part is distributed in some
momentum range. The coordinate distribution
(Fig.~\ref{Fig:distribution_x_time}a) shows periodic peaks with
the period of the lattice potential $a_x$. These peaks are
positioned  at local minima of the total potential energy, formed
by the periodic lattice potential and the potential of the applied
dc electric field (cf. Fig.~\ref{Fig:system}). It is clear that if
the dc electric field is chosen to be large enough, the local
minima of the potential energy no longer exist and the electrons
can not be captured. Once captured, such electrons do not
influence the dynamics of the moving electrons.

The electrons which are not captured by the potential minima at
$\nu_0 t=200$ are distributed already not completely homogeneously
but still quite evenly in space (see
Fig.~\ref{Fig:distribution_x_time}a). Actually, on a short time
scale, comprising however tens of oscillations periods, all of
them have the same periodic dynamics and oscillate in momentum
space between the upper and lower bounds which are strongly
pronounced in Fig.~\ref{Fig:distribution_p_time}a (and later in
Figs.~\ref{Fig:distribution_p_time}b and
\ref{Fig:distribution_p_time}c). The difference between the
electrons, leading to their distribution in space and momentum,
comes from their different phases. At the initial stage, the
phases are determined by the initial conditions and the short
relaxation stage in the beginning. The interaction between the
electrons via the radiative damping term leads to adiabatic
changes of these phases. The electrons feel each other and start
to synchronize their motion.

At $\nu_0 t=1500$ (see Fig.~\ref{Fig:distribution_p_time}b and
Fig.~\ref{Fig:distribution_x_time}b) the synchronization is
already distinguishable. We see that in both distributions moving
electrons
group together forming oscillating humps. Then between $\nu_0
t=1800$ and $\nu_0 t=2100$ (see
Figs.~\ref{Fig:distribution_p_time}c,d and
Figs.~\ref{Fig:distribution_x_time}c,d) the synchronization
becomes strongly pronounced. At $\nu_0 t=4000$ and $\nu_0 t=5000$
(see Figs.~\ref{Fig:distribution_p_time}e,f and
Figs.~\ref{Fig:distribution_x_time}e,f) all moving particles have
practically the same phase. Therefore all of them have practically
the same momentum, which oscillates between the same lower and
upper bounds. In the coordinate space the electrons cluster in
separate bunches, like in the case of the synchrotron radiation
\cite{Ternov1995}, which move through the periodic potential with
a constant and a periodically oscillating components of the
velocity. The distance between the bunches amounts to exactly the
period of the lattice potential $a_x$. The width of the bunches,
both in the momentum and coordinate spaces, tends to zero with
time. \footnote{In reality, they should possess a certain
broadening, whose description requires a more rigorous
consideration of the scattering processes, going beyond our
molecular dynamics model.} Thus, in the synchronized state the
dynamics of all the electrons is similar to the electron dynamics
in the single-electron model \cite{Mikhailov2013} but with an
additional nonlinear decay term corresponding to the radiative
damping. In the case of a relatively small radiative-damping rate
$\Gamma$, $\Gamma\ll\gamma$, the amplitude of the oscillating
current component in the synchronized state is approximately
proportional to the electron density. From
Fig.~\ref{Fig:E_S_current}b we can see that this amplitude depends
on the applied dc electric field $E_0$.

For small values of $E_0$, the electrons move so slowly that they
lose more energy due to the relaxation as they move between two
neighboring local maxima of the total potential (cf.
Fig.~\ref{Fig:system}) than the difference in energy between these
two maxima. Therefore, independent of the initial conditions,  at
some moment they can not overcome the next potential barrier and
relax to a local minimum of the potential. An increase of $E_0$
leads to a situation when two possible types of trajectories in
the phase space become possible: bound and unbound. Depending on
initial conditions, each electron can either relax to one of the
local potential minima or to propagate in space so that the energy
lost in the course of motion between two local potential maxima is
compensated by the energy difference in the height of these
maxima. The larger the value of the applied dc electric field, the
less is the number of electrons captured by the local minima of
the potential. There is a very narrow range of the dc electric
field values, where this effect leads to a rapid increase of the
oscillation amplitude of the stabilized current with $E_0$. Then,
as seen from Fig.~\ref{Fig:E_S_current}b, the amplitude of the
stationary current oscillations starts to drop with the increase
of $E_0$.
Further increase of the applied dc electric field leads to a
situation when there are no local minima of the potential and all
electrons follow unbound trajectories in the phase space. The
current oscillation amplitudes continue to decrease with the
growth of $E_0$, since the kinetic energy gained by the electrons
increases as compared to their potential energy determined by the
periodic potential. Also the synchronization time $\tau_{\rm
sync}$ increases dramatically with the increase of $E_0$. Thus,
for a realization of a coherent emission, there is a range for
optimum dc electric field values.


\begin{figure}[t!]\centering
  \includegraphics[width=\columnwidth]{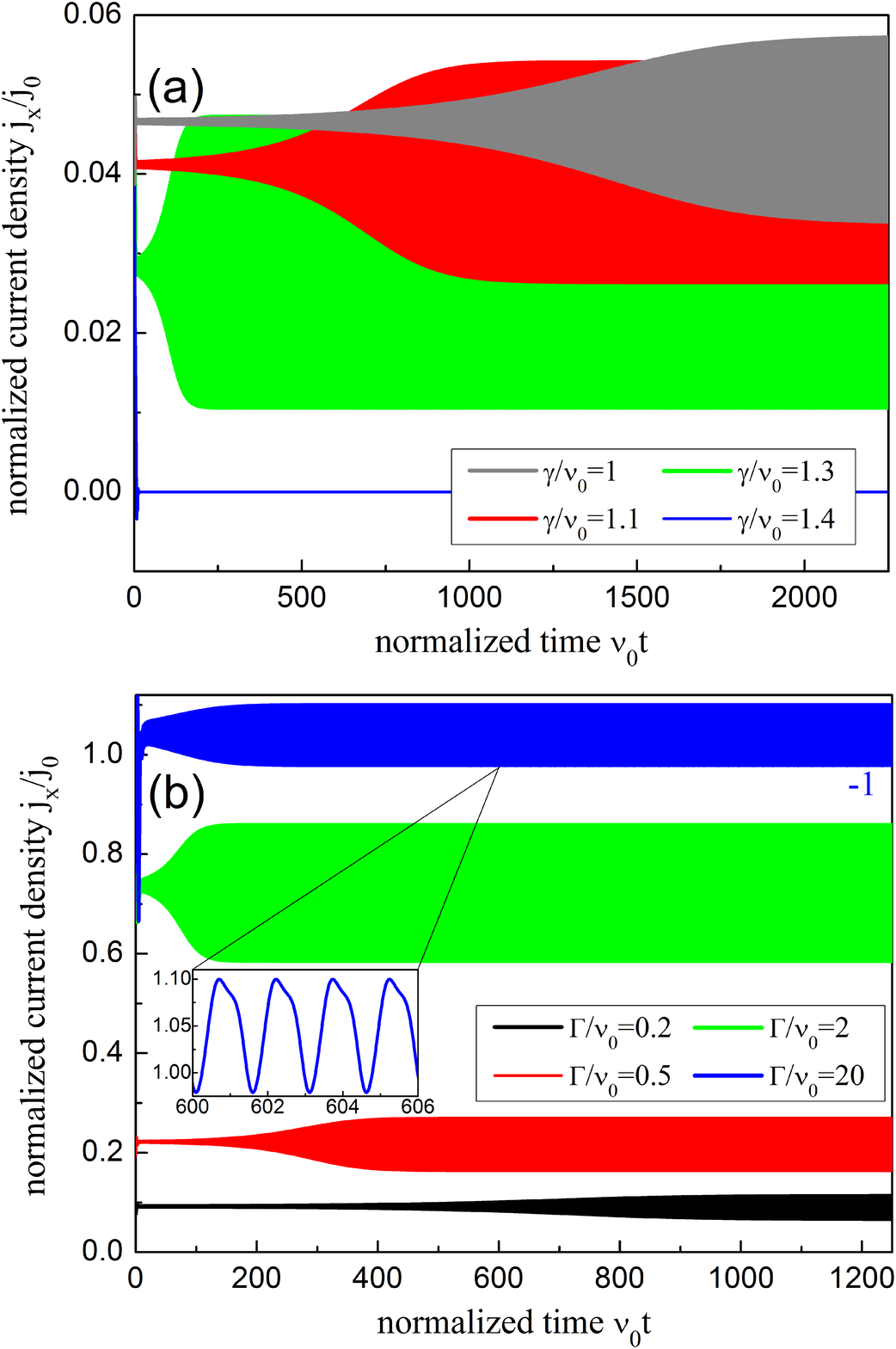}
  \caption{(Color online) (a) Current density dynamics for different values of the phenomenological
           scattering rate $\gamma$.
           Values of $\gamma$ shown in the legend increase for the curves from the top to the
           bottom of the figure.
           Other parameters of the simulations: $-eE_0a_x/\Delta_0=1$, $S=0.1$, and $\Gamma/\nu_0=0.1$.
           (b) Current density dynamics for different values of the radiative damping rate $\Gamma$.
           Values of $\Gamma$ shown in the legend increase for the curves from the bottom to the
           top of the figure.  For better readability of the figure, the upper curve for $\Gamma/\nu_0=20$
           is offset by -1 that is indicated by the corresponding number in the figure.
           Other parameters of the simulations: $-eE_0a_x/\Delta_0=1$,
           $S=0.1$, and $\gamma/\nu_0=1$. Inset shows the short-time current
           dynamics in the synchronized state for $\Gamma/\nu_0=20$.
           \label{Fig:current_g_j0}}
\end{figure}

In Fig.~\ref{Fig:current_g_j0} we illustrate the charge current
dynamics in the system at different values of  the damping
parameters $\gamma$ and $\Gamma$. Figure \ref{Fig:current_g_j0}a
shows that the influence of the scattering relaxation rate
$\gamma$ on the current oscillations is opposite to that of the dc
electric field $E_0$. This refers both to $\tau_{\rm sync}$ and to
the current oscillation amplitude. It is the ratio between $E_0$
and $\gamma$, which mostly determines the final average electron
velocity in the system, that is crucial for the system dynamics
and the properties of the generated ac current.

Figure \ref{Fig:current_g_j0}b shows that an increase of the
radiative damping rate $\Gamma$ reduces the synchronization time
$\tau_{\rm sync}$.  The time $\tau_{\rm sync}$ is roughly
inversely proportional to $\Gamma$ and saturates at
$\Gamma/\nu_0\gtrsim 1$, where we have also $\Gamma \gtrsim
\gamma$.
For small values of $\Gamma$, the radiative damping term does not
influence the single electron dynamics on a short time scale
($\Delta t\sim 1/\nu_0$) leading only to adiabatic changes of
their phase in time. The rate of these changes is proportional to
$\Gamma$. For large values of $\Gamma$ also the single electron
dynamics on a short time scale, including the initial relaxation
process, is strongly influenced by the radiative damping because
in this regime the radiative damping term is of the same order or
larger than the scattering term $-\gamma p_i$ in Eq.~\eqref{Eq:pi}

Another interesting feature observed for large values of $\Gamma$
is an appearance of a pronounced anharmonicity of the stabilized
current oscillations (see inset to Fig.~\ref{Fig:current_g_j0}b)
like in the case of the large values of the  steepness parameter
$S$ of the periodic potential (cf. Fig.~\ref{Fig:E_S_current}b).
This is again a consequence of the fact that the dynamics on a
short time scale becomes strongly affected by the radiative
damping. The anharmonicity of the charge current oscillations is
reflected in the corresponding emission spectra.

\begin{figure}[t!]\centering
  \includegraphics[width=\columnwidth]{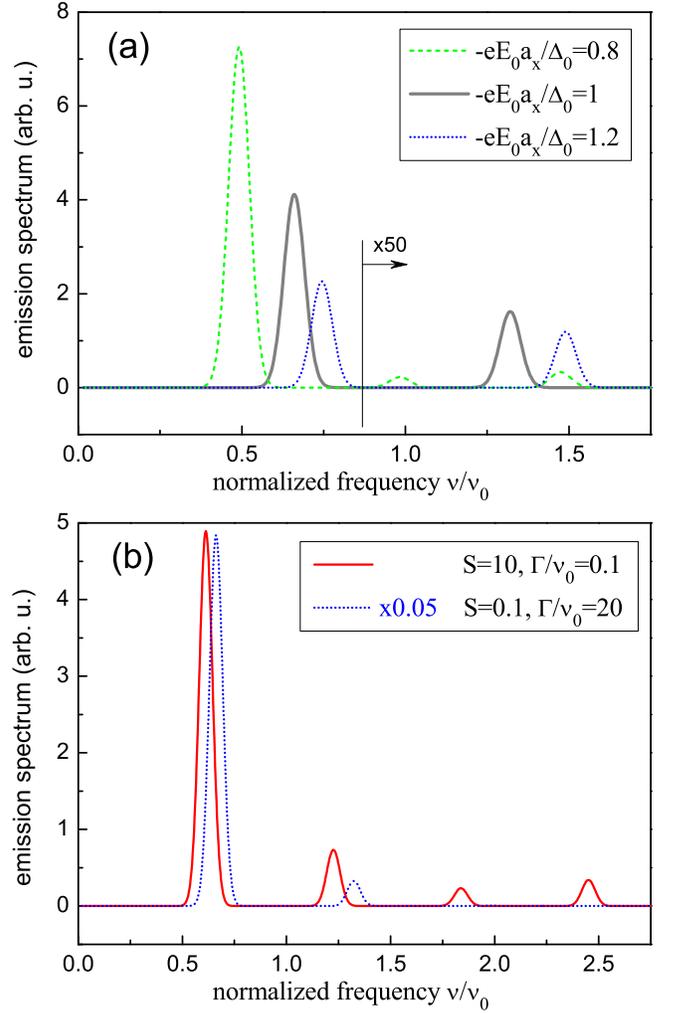}
  \caption{(Color online) (a) Emission spectra in the synchronized phase for different values of the applied dc electric field $E_0$,
  corresponding to Fig.~\ref{Fig:E_S_current}b, where $S=0.1$ To
  the right of the black vertical line the spectra are multiplied
  by the factor 50.
  (b) Emission spectra in the synchronized phase with pronounced high harmonics
  for large values of the potential steepness parameter $S$ or the radiative
  damping rate $\Gamma$, with  $-eE_0a_x/\Delta_0=1$. The spectrum for the case with $S=0.1$ and $\Gamma/\nu_0=20$ is multiplied by the factor 0.05
  (notation ''x0.05'' in the figure legend).
  For both figures a detector time window of $\Delta T\nu_0=5$ was used.\label{Fig:emission}
  }
\end{figure}

The current oscillations discussed above results in the radiation
of the electromagnetic waves. The spectra of this radiation were
calculated applying the theory of the time-dependent emission
spectra \cite{Eberly1977,Raymer1995,Moskalenko_PRA2008}, whereby a
time detection window $\Delta T$ covering several oscillation
periods was used. Figure \ref{Fig:emission} illustrates the
emission spectra calculated
in the synchronized state, when the spectra become independent of
time. In Fig.~\ref{Fig:emission}a the emission spectra are
presented for several values of the applied dc electric field
corresponding to Fig.~\ref{Fig:E_S_current}b. We see that the main
emission frequency increases with the dc electric field and is
roughly proportional to it. The main reason for this behavior is
that the main emission frequency is determined by the final
average electron velocity \cite{Mikhailov2013}, which increases
with the electric field. Analogously, larger relaxation rates
$\gamma$ lead to smaller emission frequencies (not shown here).
For small values of $S$ and $\Gamma$ the charge current
oscillations are almost harmonic. Therefore high harmonics in the
spectra are weak. Their strength rises slightly in respect to the
main harmonic when the dc electric field increases. Higher
harmonics become pronounced in case of large values of $S$, as it
was shown in the framework
of  the single-electron consideration \cite{Mikhailov2013}, or 
$\Gamma$ (see Fig.~\ref{Fig:emission}b). The latter is due to the
fact that the form of the radiative damping term [cf.
Eqs.~\eqref{Eq:pi}-\eqref{Eq:E_x_result}] is nonlinear in momentum
so that the oscillations become anharmonic when this term is large
enough and strongly influences the dynamics at the time scale
$\Delta t\sim 1/\nu_0$. Interestingly, we see from
Fig.~\ref{Fig:emission}b that a drastic increase of the radiative
damping constant $\Gamma$ leads only to a very small blue
frequency shift. This is in contrast to the influence of the
relaxation parameter $\gamma$.

\section{Conclusion}

To summarize, we have studied the radiative damping effect on the
electron dynamics in a Smith-Purcell type graphene-based device
and shown that the radiative damping not only contributes to the
electron relaxation but is also crucial for the synchronization of
their oscillatory motion and, therefore, for an effective coherent
emission of the radiation. We have illustrated this for a
graphene-based terahertz emitter. The synchronization time depends
dramatically on the ratio of the applied dc electric field to the
electron scattering rate.
Too small values of this ratio lead to
the dominance of the dissipation over the input of energy provided
by the dc electric field so that the electron oscillations decay
rapidly and the coherent emission is suppressed. However, one
should keep in mind that large values of this ratio lead to long
synchronization times and might also prohibit an effective
terahertz emission. The increase of the electron concentration
(which should be used with care in view of the relation
(\ref{thresh})) leads to the increase of the emission strength
together with the increase of the effective radiative damping
decay. For large values of the electron concentration this leads
to the saturation of the emission power and a pronounced
anharmonicity of the current oscillations, which manifests itself
through enhanced higher harmonics in the emission spectra.


\section*{Acknowledgements}

The authors thank Geoffrey Nash, Jerome Faist, Peter Kaspersen,
Isaac Luxmoore and Peter Geiser for useful discussions. The
research leading to these results has received funding from the
Deutsche Forschungsgemeinschaft and the European Community's
Seventh Framework Programme FP7/2007-2013 (project \mbox{GOSFEL},
project number 296391).

\appendix

\section{The radiative damping term in the equations of motion 
}\label{App:rad_damp}

The motion of the $i$-th electron under the action of the applied
dc field $E_0$ and the periodic potential $U(x)$ is governed by
Eq.~\eqref{Eq:pi},
where the self-consistent electric field  $E_{\rm ac}(t)$, created by all $N$ moving electrons, is related to the two-dimensional current density $j_x(t)$. 
To find this relationship, we solve the Maxwell equations
expanding the fields and currents in Fourier integral
\begin{equation}
   E_x(z,t)=\int_{-\infty}^\infty  d\omega E_x^\omega(z)e^{-i\omega t}\label{fourE}.
\end{equation}
Then the Fourier components of the electric and magnetic fields,
$E_x^\omega(z)$ and $B_y^\omega(z)$, and of the electric current $j_x^\omega$ satisfy the equations
\begin{eqnarray}
\frac{\p E_x^\omega(z)}{\p z}-i\frac \omega c
B_y^\omega(z)&=&0\;, \label{Eq:Maxw_freq_1}\\
\frac{\p B_y^\omega(z)}{\p z}-i\frac \omega c
E_x^\omega(z)&=&-\frac{4\pi}cj_x\delta(z).
\label{Eq:Maxw_freq_2}
\end{eqnarray}
The solution satisfying the conventional boundary conditions at $z=0$ and $z\to\pm\infty$ has the form
\begin{equation}
\left(
\begin{array}{c}
E_x^\omega(z)\\
B_y^\omega(z)\\
\end{array}
\right) = -\frac{2\pi}c j_x^\omega \left\{
\begin{array}{l}
\left(
\begin{array}{c}
1\\
1\\
\end{array}
\right)
e^{i\omega z/c},  \ \ \ z>0\;, \\
\left(
\begin{array}{c}
1\\
-1\\
\end{array}
\right)
e^{-i\omega z/c},  \ \ \ z<0\;. \\
\end{array}
\right. \label{Eq:E_B_solution}
\end{equation}
so that at $z=0$
\begin{equation}\label{Eq:j_solution}
  E_x^\omega(0) = -\frac{2\pi}c j_x^\omega \;.
\end{equation}
The solution \eqref{Eq:E_B_solution}--\eqref{Eq:j_solution} is
valid only at $\omega\neq 0$. If $\omega=0$, we get from
\eqref{Eq:Maxw_freq_1}--\eqref{Eq:Maxw_freq_2}
\begin{eqnarray}
  E_x^{\omega=0}(z)&=&{\rm const}\;,\label{Edcfield} \\
  B_y^{\omega=0}(z)&=&-\frac {2\pi}c
j_x^{\omega=0}\sgn(z)\;.\label{Bdcfield}
\end{eqnarray}
 The constant magnetic
field (\ref{Bdcfield}) does not influence any measured quantity
and is of no importance here. The constant dc electric field is
already taken into account in Eq.~(\ref{Eq:pi}) in the term with
$E_0$. The constant in Eq.~(\ref{Edcfield}) can therefore be taken
to be zero.

Thus we get for the Fourier component of the electric field \be
E_x^\omega(0) = -\frac{2\pi}c j_x^\omega \times
\left\{\begin{array}{c}
1, \ \ \ {\rm if}\  \omega\neq 0\;,\\
0, \ \ \ {\rm if}\  \omega= 0\;.\\
\end{array}\right. \label{Exfield}
\ee
In this form, there is no dc-current contribution to the
radiative damping, as it also should be. However, the field
(\ref{Exfield}) is discontinuous in one point which leads to a problem. Indeed, substituting \eqref{Exfield} into \eqref{fourE} and calculating the time dependent field $E_x(0,t)$ we see that
the information about $E_x^{\omega=0}(0)=0$ is lost.

This problem has arisen since we  consider an idealized situation of an infinitely long
sample. In reality the length of the sample $L$ (in the
$x$-direction) is finite, and the Fourier component of the ac
electric field vanishes not at the only one point $\omega=0$, Eq.
(\ref{Exfield}), but effectively in the range
$|\omega|\lesssim \gamma_L$, where $\gamma_L$ is given by
Eq.~\eqref{Eq:omega_L}. Therefore, for a real (finite)
system we assume the following functional dependence of $E_x^{\omega}(0)$ on the
Fourier component of the current:
\begin{equation}
  E_x^\omega(0) = -\frac{2\pi}c j_x^\omega\left[1- r(\omega)\right].
  \label{ExfieldModel}
\end{equation}
The function $r(\omega)$
should be equal to 1 at $\omega=0$, sufficiently fast tend to zero
at $|\omega|\gtrsim \gamma_L$, and should satisfy the causality
relation between the charge current and the generated ac electric
field in the time domain. As a model expression we take the
function
\begin{equation}\label{Eq:r_omega}
  r(\omega)=\frac{1}{1-i\omega/\gamma_L}
\end{equation}
with a pole in the lower complex half-plane to satisfy the causality condition.
Now we can return back to Eq.~\eqref{Eq:pi} and to calculate the
field $E_x(t)$. Performing the Fourier transform \eqref{fourE} of
the function \eqref{ExfieldModel} with the $r(\omega)$ function
from (\ref{Eq:r_omega}) we get Eq. (\ref{Eq:E_x_result}) with
$\bar j_x(t)$ from (\ref{Eq:j_bar}). Eq.~\eqref{Eq:j_bar} can be
rewritten as an ordinary differential equation
\begin{equation}\label{Eq:j_bar_ODE}
  \frac{{\rm d}\bar{j}_x}{{\rm d}t}=\omega_L(j_x-\bar{j}_x).
\end{equation}

\bibliography{report,additrefs}

\begin{thebibliography}{23}%
\makeatletter
\providecommand \@ifxundefined [1]{%
 \@ifx{#1\undefined}
}%
\providecommand \@ifnum [1]{%
 \ifnum #1\expandafter \@firstoftwo
 \else \expandafter \@secondoftwo
 \fi
}%
\providecommand \@ifx [1]{%
 \ifx #1\expandafter \@firstoftwo
 \else \expandafter \@secondoftwo
 \fi
}%
\providecommand \natexlab [1]{#1}%
\providecommand \enquote  [1]{``#1''}%
\providecommand \bibnamefont  [1]{#1}%
\providecommand \bibfnamefont [1]{#1}%
\providecommand \citenamefont [1]{#1}%
\providecommand \href@noop [0]{\@secondoftwo}%
\providecommand \href [0]{\begingroup \@sanitize@url \@href}%
\providecommand \@href[1]{\@@startlink{#1}\@@href}%
\providecommand \@@href[1]{\endgroup#1\@@endlink}%
\providecommand \@sanitize@url [0]{\catcode `\\12\catcode `\$12\catcode
  `\&12\catcode `\#12\catcode `\^12\catcode `\_12\catcode `\%12\relax}%
\providecommand \@@startlink[1]{}%
\providecommand \@@endlink[0]{}%
\providecommand \url  [0]{\begingroup\@sanitize@url \@url }%
\providecommand \@url [1]{\endgroup\@href {#1}{\urlprefix }}%
\providecommand \urlprefix  [0]{URL }%
\providecommand \Eprint [0]{\href }%
\providecommand \doibase [0]{http://dx.doi.org/}%
\providecommand \selectlanguage [0]{\@gobble}%
\providecommand \bibinfo  [0]{\@secondoftwo}%
\providecommand \bibfield  [0]{\@secondoftwo}%
\providecommand \translation [1]{[#1]}%
\providecommand \BibitemOpen [0]{}%
\providecommand \bibitemStop [0]{}%
\providecommand \bibitemNoStop [0]{.\EOS\space}%
\providecommand \EOS [0]{\spacefactor3000\relax}%
\providecommand \BibitemShut  [1]{\csname bibitem#1\endcsname}%
\let\auto@bib@innerbib\@empty
\bibitem [{\citenamefont {Smith}\ and\ \citenamefont
  {Purcell}(1953)}]{Smith1953}%
  \BibitemOpen
  \bibfield  {author} {\bibinfo {author} {\bibfnamefont {S.~J.}\ \bibnamefont
  {Smith}}\ and\ \bibinfo {author} {\bibfnamefont {E.~M.}\ \bibnamefont
  {Purcell}},\ }\href@noop {} {\bibfield  {journal} {\bibinfo  {journal} {Phys.
  Rev.}\ }\textbf {\bibinfo {volume} {92}},\ \bibinfo {pages} {1069} (\bibinfo
  {year} {1953})}\BibitemShut {NoStop}%
\bibitem [{\citenamefont {Tsui}\ \emph {et~al.}(1980)\citenamefont {Tsui},
  \citenamefont {Gornik},\ and\ \citenamefont {Logan}}]{Tsui80}%
  \BibitemOpen
  \bibfield  {author} {\bibinfo {author} {\bibfnamefont {D.~C.}\ \bibnamefont
  {Tsui}}, \bibinfo {author} {\bibfnamefont {E.}~\bibnamefont {Gornik}}, \ and\
  \bibinfo {author} {\bibfnamefont {R.~A.}\ \bibnamefont {Logan}},\ }\href@noop
  {} {\bibfield  {journal} {\bibinfo  {journal} {Solid State Commun.}\ }\textbf
  {\bibinfo {volume} {35}},\ \bibinfo {pages} {875} (\bibinfo {year}
  {1980})}\BibitemShut {NoStop}%
\bibitem [{\citenamefont {Hirakawa}\ \emph {et~al.}(1995)\citenamefont
  {Hirakawa}, \citenamefont {Yamanaka}, \citenamefont {Grayson},\ and\
  \citenamefont {Tsui}}]{Hirakawa95}%
  \BibitemOpen
  \bibfield  {author} {\bibinfo {author} {\bibfnamefont {K.}~\bibnamefont
  {Hirakawa}}, \bibinfo {author} {\bibfnamefont {K.}~\bibnamefont {Yamanaka}},
  \bibinfo {author} {\bibfnamefont {M.}~\bibnamefont {Grayson}}, \ and\
  \bibinfo {author} {\bibfnamefont {D.~C.}\ \bibnamefont {Tsui}},\ }\href@noop
  {} {\bibfield  {journal} {\bibinfo  {journal} {Appl. Phys. Lett.}\ }\textbf
  {\bibinfo {volume} {67}},\ \bibinfo {pages} {2326} (\bibinfo {year}
  {1995})}\BibitemShut {NoStop}%
\bibitem [{\citenamefont {Otsuji}\ \emph {et~al.}(2011)\citenamefont {Otsuji},
  \citenamefont {Watanabe}, \citenamefont {E.}, \citenamefont {H.},
  \citenamefont {Komori}, \citenamefont {Satou}, \citenamefont {Suemitsu},
  \citenamefont {Suemitsu}, \citenamefont {Sano}, \citenamefont {Knap},\ and\
  \citenamefont {Ryzhii}}]{Otsuji11}%
  \BibitemOpen
  \bibfield  {author} {\bibinfo {author} {\bibfnamefont {T.}~\bibnamefont
  {Otsuji}}, \bibinfo {author} {\bibfnamefont {T.}~\bibnamefont {Watanabe}},
  \bibinfo {author} {\bibfnamefont {M.~A.}\ \bibnamefont {E.}}, \bibinfo
  {author} {\bibfnamefont {K.}~\bibnamefont {H.}}, \bibinfo {author}
  {\bibfnamefont {T.}~\bibnamefont {Komori}}, \bibinfo {author} {\bibfnamefont
  {A.}~\bibnamefont {Satou}}, \bibinfo {author} {\bibfnamefont
  {T.}~\bibnamefont {Suemitsu}}, \bibinfo {author} {\bibfnamefont
  {M.}~\bibnamefont {Suemitsu}}, \bibinfo {author} {\bibfnamefont
  {E.}~\bibnamefont {Sano}}, \bibinfo {author} {\bibfnamefont {W.}~\bibnamefont
  {Knap}}, \ and\ \bibinfo {author} {\bibfnamefont {V.}~\bibnamefont
  {Ryzhii}},\ }\href@noop {} {\bibfield  {journal} {\bibinfo  {journal} {J.
  Infrared Milli. Terahz. Waves}\ }\textbf {\bibinfo {volume} {32}},\ \bibinfo
  {pages} {629} (\bibinfo {year} {2011})}\BibitemShut {NoStop}%
\bibitem [{\citenamefont {Mikhailov}(1998)}]{Mikhailov1998}%
  \BibitemOpen
  \bibfield  {author} {\bibinfo {author} {\bibfnamefont {S.~A.}\ \bibnamefont
  {Mikhailov}},\ }\href@noop {} {\bibfield  {journal} {\bibinfo  {journal}
  {Phys. Rev. B}\ }\textbf {\bibinfo {volume} {58}},\ \bibinfo {pages} {1517}
  (\bibinfo {year} {1998})}\BibitemShut {NoStop}%
\bibitem [{\citenamefont {Novoselov}\ \emph {et~al.}(2004)\citenamefont
  {Novoselov}, \citenamefont {Geim}, \citenamefont {Morozov}, \citenamefont
  {Jiang}, \citenamefont {Zhang}, \citenamefont {Dubonos}, \citenamefont
  {Grigorieva},\ and\ \citenamefont {Firsov}}]{Novoselov04}%
  \BibitemOpen
  \bibfield  {author} {\bibinfo {author} {\bibfnamefont {K.~S.}\ \bibnamefont
  {Novoselov}}, \bibinfo {author} {\bibfnamefont {A.~K.}\ \bibnamefont {Geim}},
  \bibinfo {author} {\bibfnamefont {S.~V.}\ \bibnamefont {Morozov}}, \bibinfo
  {author} {\bibfnamefont {D.}~\bibnamefont {Jiang}}, \bibinfo {author}
  {\bibfnamefont {Y.}~\bibnamefont {Zhang}}, \bibinfo {author} {\bibfnamefont
  {S.~V.}\ \bibnamefont {Dubonos}}, \bibinfo {author} {\bibfnamefont {I.~V.}\
  \bibnamefont {Grigorieva}}, \ and\ \bibinfo {author} {\bibfnamefont {A.~A.}\
  \bibnamefont {Firsov}},\ }\href@noop {} {\bibfield  {journal} {\bibinfo
  {journal} {Science}\ }\textbf {\bibinfo {volume} {306}},\ \bibinfo {pages}
  {666} (\bibinfo {year} {2004})}\BibitemShut {NoStop}%
\bibitem [{\citenamefont {Novoselov}\ \emph {et~al.}(2005)\citenamefont
  {Novoselov}, \citenamefont {Geim}, \citenamefont {Morozov}, \citenamefont
  {Jiang}, \citenamefont {Katsnelson}, \citenamefont {Grigorieva},
  \citenamefont {Dubonos},\ and\ \citenamefont {Firsov}}]{Novoselov05}%
  \BibitemOpen
  \bibfield  {author} {\bibinfo {author} {\bibfnamefont {K.~S.}\ \bibnamefont
  {Novoselov}}, \bibinfo {author} {\bibfnamefont {A.~K.}\ \bibnamefont {Geim}},
  \bibinfo {author} {\bibfnamefont {S.~V.}\ \bibnamefont {Morozov}}, \bibinfo
  {author} {\bibfnamefont {D.}~\bibnamefont {Jiang}}, \bibinfo {author}
  {\bibfnamefont {M.~I.}\ \bibnamefont {Katsnelson}}, \bibinfo {author}
  {\bibfnamefont {I.~V.}\ \bibnamefont {Grigorieva}}, \bibinfo {author}
  {\bibfnamefont {S.~V.}\ \bibnamefont {Dubonos}}, \ and\ \bibinfo {author}
  {\bibfnamefont {A.~A.}\ \bibnamefont {Firsov}},\ }\href@noop {} {\bibfield
  {journal} {\bibinfo  {journal} {Nature}\ }\textbf {\bibinfo {volume} {438}},\
  \bibinfo {pages} {197} (\bibinfo {year} {2005})}\BibitemShut {NoStop}%
\bibitem [{\citenamefont {Zhang}\ \emph {et~al.}(2005)\citenamefont {Zhang},
  \citenamefont {Tan}, \citenamefont {Stormer},\ and\ \citenamefont
  {Kim}}]{Zhang05}%
  \BibitemOpen
  \bibfield  {author} {\bibinfo {author} {\bibfnamefont {Y.}~\bibnamefont
  {Zhang}}, \bibinfo {author} {\bibfnamefont {Y.-W.}\ \bibnamefont {Tan}},
  \bibinfo {author} {\bibfnamefont {H.~L.}\ \bibnamefont {Stormer}}, \ and\
  \bibinfo {author} {\bibfnamefont {P.}~\bibnamefont {Kim}},\ }\href@noop {}
  {\bibfield  {journal} {\bibinfo  {journal} {Nature}\ }\textbf {\bibinfo
  {volume} {438}},\ \bibinfo {pages} {201} (\bibinfo {year}
  {2005})}\BibitemShut {NoStop}%
\bibitem [{\citenamefont {Mikhailov}(2013)}]{Mikhailov2013}%
  \BibitemOpen
  \bibfield  {author} {\bibinfo {author} {\bibfnamefont {S.~A.}\ \bibnamefont
  {Mikhailov}},\ }\href@noop {} {\bibfield  {journal} {\bibinfo  {journal}
  {Phys. Rev. B}\ }\textbf {\bibinfo {volume} {87}},\ \bibinfo {pages} {115405}
  (\bibinfo {year} {2013})}\BibitemShut {NoStop}%
\bibitem [{\citenamefont {Strogatz}(2001)}]{Strogatz2001}%
  \BibitemOpen
  \bibfield  {author} {\bibinfo {author} {\bibfnamefont {S.~H.}\ \bibnamefont
  {Strogatz}},\ }\href@noop {} {\emph {\bibinfo {title} {Nonlinear Dynamics And
  Chaos: With Applications To Physics, Biology, Chemistry, And Engineering}}}\
  (\bibinfo  {publisher} {Westview Press, Boulder},\ \bibinfo {year}
  {2001})\BibitemShut {NoStop}%
\bibitem [{\citenamefont {Pikovsky}\ \emph {et~al.}(2003)\citenamefont
  {Pikovsky}, \citenamefont {Rosenblum},\ and\ \citenamefont
  {Kurths}}]{Pikovsky2003}%
  \BibitemOpen
  \bibfield  {author} {\bibinfo {author} {\bibfnamefont {A.}~\bibnamefont
  {Pikovsky}}, \bibinfo {author} {\bibfnamefont {M.}~\bibnamefont {Rosenblum}},
  \ and\ \bibinfo {author} {\bibfnamefont {J.}~\bibnamefont {Kurths}},\
  }\href@noop {} {\emph {\bibinfo {title} {Synchronization: A Universal Concept
  in Nonlinear Sciences}}},\ Cambridge Nonlinear Science Series, Cambridge\
  (\bibinfo  {publisher} {Cambridge University Press},\ \bibinfo {year}
  {2003})\BibitemShut {NoStop}%
\bibitem [{\citenamefont {Arenas}\ \emph {et~al.}(2008)\citenamefont {Arenas},
  \citenamefont {Diaz-Guilera}, \citenamefont {Kurths}, \citenamefont
  {Moreno},\ and\ \citenamefont {Zhou}}]{Arenas2008}%
  \BibitemOpen
  \bibfield  {author} {\bibinfo {author} {\bibfnamefont {A.}~\bibnamefont
  {Arenas}}, \bibinfo {author} {\bibfnamefont {A.}~\bibnamefont
  {Diaz-Guilera}}, \bibinfo {author} {\bibfnamefont {J.}~\bibnamefont
  {Kurths}}, \bibinfo {author} {\bibfnamefont {Y.}~\bibnamefont {Moreno}}, \
  and\ \bibinfo {author} {\bibfnamefont {C.}~\bibnamefont {Zhou}},\ }\href@noop
  {} {\bibfield  {journal} {\bibinfo  {journal} {Phys. Rep.}\ }\textbf
  {\bibinfo {volume} {469}},\ \bibinfo {pages} {93 } (\bibinfo {year}
  {2008})}\BibitemShut {NoStop}%
\bibitem [{\citenamefont {Strogatz}(2000)}]{Strogatz2000}%
  \BibitemOpen
  \bibfield  {author} {\bibinfo {author} {\bibfnamefont {S.~H.}\ \bibnamefont
  {Strogatz}},\ }\href@noop {} {\bibfield  {journal} {\bibinfo  {journal}
  {Phys. D}\ }\textbf {\bibinfo {volume} {143}},\ \bibinfo {pages} {1 }
  (\bibinfo {year} {2000})}\BibitemShut {NoStop}%
\bibitem [{\citenamefont {Brey}\ and\ \citenamefont {Fertig}(2006)}]{Brey2006}%
  \BibitemOpen
  \bibfield  {author} {\bibinfo {author} {\bibfnamefont {L.}~\bibnamefont
  {Brey}}\ and\ \bibinfo {author} {\bibfnamefont {H.~A.}\ \bibnamefont
  {Fertig}},\ }\href@noop {} {\bibfield  {journal} {\bibinfo  {journal} {Phys.
  Rev. B}\ }\textbf {\bibinfo {volume} {73}},\ \bibinfo {pages} {235411}
  (\bibinfo {year} {2006})}\BibitemShut {NoStop}%
\bibitem [{\citenamefont {Yang}\ \emph {et~al.}(2007)\citenamefont {Yang},
  \citenamefont {Park}, \citenamefont {Son}, \citenamefont {Cohen},\ and\
  \citenamefont {Louie}}]{Yang2007}%
  \BibitemOpen
  \bibfield  {author} {\bibinfo {author} {\bibfnamefont {L.}~\bibnamefont
  {Yang}}, \bibinfo {author} {\bibfnamefont {C.-H.}\ \bibnamefont {Park}},
  \bibinfo {author} {\bibfnamefont {Y.-W.}\ \bibnamefont {Son}}, \bibinfo
  {author} {\bibfnamefont {M.~L.}\ \bibnamefont {Cohen}}, \ and\ \bibinfo
  {author} {\bibfnamefont {S.~G.}\ \bibnamefont {Louie}},\ }\href@noop {}
  {\bibfield  {journal} {\bibinfo  {journal} {Phys. Rev. Lett.}\ }\textbf
  {\bibinfo {volume} {99}},\ \bibinfo {pages} {186801} (\bibinfo {year}
  {2007})}\BibitemShut {NoStop}%
\bibitem [{\citenamefont {Akhmerov}\ and\ \citenamefont
  {Beenakker}(2008)}]{Akhmerov2008}%
  \BibitemOpen
  \bibfield  {author} {\bibinfo {author} {\bibfnamefont {A.~R.}\ \bibnamefont
  {Akhmerov}}\ and\ \bibinfo {author} {\bibfnamefont {C.~W.~J.}\ \bibnamefont
  {Beenakker}},\ }\href@noop {} {\bibfield  {journal} {\bibinfo  {journal}
  {Phys. Rev. B}\ }\textbf {\bibinfo {volume} {77}},\ \bibinfo {pages} {085423}
  (\bibinfo {year} {2008})}\BibitemShut {NoStop}%
\bibitem [{\citenamefont {Mikhailov}\ and\ \citenamefont
  {Ziegler}(2008)}]{Mikhailov08a}%
  \BibitemOpen
  \bibfield  {author} {\bibinfo {author} {\bibfnamefont {S.~A.}\ \bibnamefont
  {Mikhailov}}\ and\ \bibinfo {author} {\bibfnamefont {K.}~\bibnamefont
  {Ziegler}},\ }\href@noop {} {\bibfield  {journal} {\bibinfo  {journal} {J.
  Phys. Condens. Matter}\ }\textbf {\bibinfo {volume} {20}},\ \bibinfo {pages}
  {384204} (\bibinfo {year} {2008})}\BibitemShut {NoStop}%
\bibitem [{\citenamefont {Krumbholz}\ \emph {et~al.}(2006)\citenamefont
  {Krumbholz}, \citenamefont {Gerlach}, \citenamefont {Rutz}, \citenamefont
  {Koch}, \citenamefont {Piesiewicz}, \citenamefont {K{\"{u}}rner},\ and\
  \citenamefont {Mittleman}}]{Krumbholz2006}%
  \BibitemOpen
  \bibfield  {author} {\bibinfo {author} {\bibfnamefont {N.}~\bibnamefont
  {Krumbholz}}, \bibinfo {author} {\bibfnamefont {K.}~\bibnamefont {Gerlach}},
  \bibinfo {author} {\bibfnamefont {F.}~\bibnamefont {Rutz}}, \bibinfo {author}
  {\bibfnamefont {M.}~\bibnamefont {Koch}}, \bibinfo {author} {\bibfnamefont
  {R.}~\bibnamefont {Piesiewicz}}, \bibinfo {author} {\bibfnamefont
  {T.}~\bibnamefont {K{\"{u}}rner}}, \ and\ \bibinfo {author} {\bibfnamefont
  {D.}~\bibnamefont {Mittleman}},\ }\href@noop {} {\bibfield  {journal}
  {\bibinfo  {journal} {Appl. Phys. Lett.}\ }\textbf {\bibinfo {volume} {88}},\
  \bibinfo {eid} {202905} (\bibinfo {year} {2006})}\BibitemShut {NoStop}%
\bibitem [{\citenamefont {Ternov}(1995)}]{Ternov1995}%
  \BibitemOpen
  \bibfield  {author} {\bibinfo {author} {\bibfnamefont {I.~M.}\ \bibnamefont
  {Ternov}},\ }\href@noop {} {\bibfield  {journal} {\bibinfo  {journal}
  {Phys.-Usp.}\ }\textbf {\bibinfo {volume} {38}},\ \bibinfo {pages} {409}
  (\bibinfo {year} {1995})}\BibitemShut {NoStop}%
\bibitem [{Note1()}]{Note1}%
  \BibitemOpen
  \bibinfo {note} {In reality, they should possess a certain broadening, whose
  description requires a more rigorous consideration of the scattering
  processes, going beyond our molecular dynamics model.}\BibitemShut {Stop}%
\bibitem [{\citenamefont {Eberly}\ and\ \citenamefont
  {W{\'o}dkiewicz}(1977)}]{Eberly1977}%
  \BibitemOpen
  \bibfield  {author} {\bibinfo {author} {\bibfnamefont {J.~H.}\ \bibnamefont
  {Eberly}}\ and\ \bibinfo {author} {\bibfnamefont {K.}~\bibnamefont
  {W{\'o}dkiewicz}},\ }\href@noop {} {\bibfield  {journal} {\bibinfo  {journal}
  {J. Opt. Soc. Am.}\ }\textbf {\bibinfo {volume} {67}},\ \bibinfo {pages}
  {1252} (\bibinfo {year} {1977})}\BibitemShut {NoStop}%
\bibitem [{\citenamefont {Raymer}\ \emph {et~al.}(1997)\citenamefont {Raymer},
  \citenamefont {Cooper}, \citenamefont {Carmichael}, \citenamefont {Beck},\
  and\ \citenamefont {Smithey}}]{Raymer1995}%
  \BibitemOpen
  \bibfield  {author} {\bibinfo {author} {\bibfnamefont {M.~G.}\ \bibnamefont
  {Raymer}}, \bibinfo {author} {\bibfnamefont {J.}~\bibnamefont {Cooper}},
  \bibinfo {author} {\bibfnamefont {H.~J.}\ \bibnamefont {Carmichael}},
  \bibinfo {author} {\bibfnamefont {M.}~\bibnamefont {Beck}}, \ and\ \bibinfo
  {author} {\bibfnamefont {D.~T.}\ \bibnamefont {Smithey}},\ }\href@noop {}
  {\bibfield  {journal} {\bibinfo  {journal} {J. Opt. Soc. Am. B}\ }\textbf
  {\bibinfo {volume} {12}},\ \bibinfo {pages} {1801} (\bibinfo {year}
  {1997})}\BibitemShut {NoStop}%
\bibitem [{\citenamefont {Moskalenko}\ and\ \citenamefont
  {Berakdar}(2008)}]{Moskalenko_PRA2008}%
  \BibitemOpen
  \bibfield  {author} {\bibinfo {author} {\bibfnamefont {A.~S.}\ \bibnamefont
  {Moskalenko}}\ and\ \bibinfo {author} {\bibfnamefont {J.}~\bibnamefont
  {Berakdar}},\ }\href@noop {} {\bibfield  {journal} {\bibinfo  {journal}
  {Phys. Rev. A}\ }\textbf {\bibinfo {volume} {78}},\ \bibinfo {pages}
  {051804(R)} (\bibinfo {year} {2008})}\BibitemShut {NoStop}%
\end{thebibliography}%

\end{document}